\documentclass[letterpaper]{article}
\usepackage[T1]{fontenc}
\usepackage{geometry}
\usepackage{setspace}
\usepackage[backend=biber, style=chem-acs, maxnames=99,minnames=99]{biblatex}
\usepackage{graphicx}
\usepackage{float}
\newfloat{scheme}{htbp}{los}
\floatname{scheme}{Scheme}
\floatname{chart}{Chart}
\newfloat{graph}{htbp}{loh}
\usepackage[version = 4]{mhchem} 
\usepackage[colorlinks=true, linkcolor=black, citecolor=black, urlcolor=blue]{hyperref}
\usepackage{epstopdf}
\usepackage{bm,amsmath,amssymb} 
\usepackage{wasysym}
\usepackage{color, soul} 
\usepackage{dcolumn}   
\usepackage{multirow}
\usepackage{booktabs}
\usepackage{makecell} 
\usepackage{lineno}
\usepackage{authblk}
\newcommand{\etal}{{\em et al}.\ }
 
\newcommand{\note}[1]{{\color{black}{{#1}}}}
\newcommand{\notes}[1]{{\color{black}{{#1}}}} 

\author[1,$\dagger$]{Denan Li}
\author[1,$\dagger$]{Jiyuan Yang}
\author[1,2]{Xiangkai Chen}
\author[1]{Lintao Yu}
\author[1,2,*]{Shi Liu}
\affil[1]{Department of Physics, School of Science, Westlake University, Hangzhou, Zhejiang 310030, China}
\affil[2]{Institute of Natural Sciences, Westlake Institute for Advanced Study, Hangzhou, Zhejiang 310024, China}
\affil[$\dagger$]{These authors contributed equally}
\affil[*]{Corresponding author: liushi@westlake.edu.cn}
\title{\note{Are Foundational Atomistic Models Reliable for Finite-Temperature Molecular Dynamics?}}

\begin{document}
\maketitle
\newpage
\begin{abstract}
\note{Machine learning force fields have emerged as promising tools for molecular dynamics (MD) simulations, potentially offering quantum-mechanical accuracy with the efficiency of classical MD. Inspired by foundational large language models, recent years have seen considerable progress in developing foundational atomistic models, sometimes referred to as universal force fields, designed to cover most elements in the periodic table. This Perspective adopts a practitioner's viewpoint to ask a critical question: Are these foundational atomistic models reliable for one of their most compelling applications, in particular simulating finite-temperature dynamics? Instead of a broad benchmark, we use the canonical ferroelectric-paraelectric phase transition in PbTiO$_3$ as a focused case study to evaluate prominent foundational atomistic models. Our findings suggest a potential disconnect between static accuracy and dynamic reliability. While 0 K properties are often well-reproduced, we observed that the models can struggle to consistently capture the correct phase transition, sometimes exhibiting simulation instabilities. We believe these challenges may stem from inherent biases in training data and a limited description of anharmonicity. These observed shortcomings, though demonstrated on a single system, appear to point to broader, systemic challenges that can be addressed with targeted fine-tuning. This Perspective serves not to rank models, but to initiate a crucial discussion on the practical readiness of foundational atomistic models and to explore future directions for their improvement.}
\end{abstract}
\newpage

\note{\section{Introduction}}
Artificial intelligence (AI) is rapidly emerging as the fifth paradigm of scientific research, joining the established paradigms of experiments, theory, computation, and data. This transformative technology is fundamentally reshaping the nature of scientific inquiry and has the potential to significantly accelerate the pace of scientific discovery. The recognition of AI's contributions to science, highlighted by its acknowledgment in the 2024 Nobel Prizes in both Physics and Chemistry\cite{nobel2024physics,nobel2024chemistry}, firmly establishes the era of ``AI for Science." In materials science, AI holds immense potential to elucidate complex structure-property relationships, thereby enhancing and expediting the processes of materials discovery and design. 

In this Perspective, we primarily focus on machine learning force fields (MLFFs) for classical MD simulations. Classical MD simulations employ parameterized interatomic potentials, enabling computationally efficient exploration of dynamic processes across large temporal and spatial scales. These simulations not only reveal atomic-level mechanisms but also provide foundational data for coarse-grained models~\cite{Schilling22p1,Sun21p645527}, further extending their importance in multiscale simulations. Classical MD simulations have long been an indispensable tool for computer-aided drug design~\cite{Macalino15p1686}, driven by classical force fields that accurately describe biomolecular interactions in proteins and nucleic acids~\cite{Liebl23p2841,Samuel23p9863}.
In contrast, the adoption of MD in computer-aided materials discovery lags behind due to the absence of force fields capable of handling diverse elements, especially transition-metal oxides and complex alloys. This challenge mainly stems from the high-dimensional potential energy surface inherent in multielement systems, where traditional analytical functionals struggle to balance accuracy and generality.

The emergence of MLFFs is transforming the field of MD simulations. By leveraging advanced techniques such as deep neural networks and graph neural networks, MLFFs could achieve quantum-mechanical accuracy while retaining the computational efficiency of classical MD. The standard protocol for developing an MLFF involves training the model on databases computed using density functional theory (DFT), which include energies, atomic forces, and virial tensors across a diverse set of atomic configurations.
\note{A significant recent trend in materials modeling, paralleling the rise of foundational models in machine learning, is the development of what are often called ``universal force fields" ~\cite{Chen22p718,Deng23p1031,Batatia23parXivv1,Xie24p3525,Neumann24parXivv1,Zhang24p293v1,Kim24p1042,Yin25parXivv1,Yang24mattersimv1}. In this Perspective, we refer to these as \textbf{foundational atomistic models}, a term we find useful to highlight their intended role as a general-purpose base for a wide range of downstream applications. These models are characterized by their training on vast and chemically diverse datasets, often encompassing a large portion of the periodic table, with the ultimate aim of enabling efficient simulations of complex materials systems at an accuracy approaching that of first-principles methods.}
Notable examples include M3GNet~\cite{Chen22p718}, CHGNet~\cite{Deng23p1031}, and MACE~\cite{Batatia23parXivv1}, which are based on graph neural network architectures and are trained on extensive, materials science databases~\cite{Jain13p1v1,Schmidt22p64,Barroso24parXivv1}. Proprietary advancements include GNoME~\cite{Merchant23p80}, built upon E(3)-equivariant graph neural networks, and PFP~\cite{Takamoto22p1}, which leverages the TeaNet architecture~\cite{Takamoto22p111280} to combine attention mechanisms with graph-based atomic representations. 
The GPTFF model~\cite{Xie24p3525} integrates graph neural network and transformer architectures with attention mechanisms, is trained on the proprietary Atomly database. The DPA-2 model~\cite{Zhang24p293v1} positions itself as a pre-trained model covering more than 90 elements. It is designed to significantly reduce downstream data requirements by leveraging transfer learning, enabling efficient on-demand fine-tuning to create tailored models for specific materials of user interest. These developments mark a paradigm shift toward general-purpose force fields capable of simulating complex multielement systems, from battery electrolytes to high-entropy alloys. Recently,  Riebesell~\etal~developed \texttt{Matbench Discovery}, an evaluation framework
for MLFFs, applied as pre-filters for high-throughput searches of stable inorganic crystals~\cite{riebesell25p836}.


\note{Many excellent reviews have focused on comprehensively benchmarking these foundational atomistic models, often ranking them based on their accuracy for static properties across vast chemical spaces~\cite{wines25p2105,yu2024pe58,kim23p51434,mortazavi25p2403876,deng25p9,Poltavsky25p3738}. In this Perspective, we adopt a deliberately narrower, more practitioner-focused viewpoint. The distinctive advantage of MD over static calculations is its ability to reveal time-dependent atomic behaviors and emergent properties. From our perspective, this suggests that while static accuracy is a necessary foundation, a critical test of a foundational atomistic model's practical utility is its performance under realistic, dynamic conditions. This leads us to pose a question that we believe precedes simple model selection: Can a practitioner trust an ``out-of-the-box" foundational atomistic model to reliably capture the complex physics of their specific material system, particularly during finite-temperature MD simulations? Therefore, our focus here is not on static accuracy alone, but on the ability of these models to reproduce the dynamic behavior of materials over the timescales, particularly tens to hundreds of picoseconds, where emergent phenomena like phase transitions occur.}

\notes{
\section{Methods}
All calculations are performed using the ASE~\cite{Larsen17p273002}, with each MLFF integrated as an ASE calculator to compute energy, forces, and stress. To simulate the temperature-driven phase transition of $\mathrm{PbTiO_3}$, a $5\times5\times5$ supercell containing 625 atoms is constructed from the ground-state structure. MD simulations are carried out in the $NPT$ ensemble using a Parrinello–Rahman barostat coupled with a Nos\'e–Hoover thermostat. At each temperature, the simulation runs with a 2 fs timestep for 50,000 steps, totaling 100 ps. The last 50 ps of the trajectory is used to compute the averaged lattice constants. Our tests confirm that the cumulative average of the $c/a$ ratio converges after a 10-ps production trajectory.
 For the $NVT$ ensemble simulation, the same $5\times5\times5$ supercell is used, with lattice parameters fixed at their experimental values measured at room temperature. Langevin dynamics are employed with a 2 fs timestep for 50,000 steps. For performance benchmarking, all computations are performed on a single V100-SXM2-16G GPU, with each data point averaged over three independent runs.  
}

\note{\section{Benchmark Design}}
In the following (admittedly not comprehensive) performance assessment, we use the temperature-driven ferroelectric-paraelectric phase transition of PbTiO$_3$ as a test case, referred to as the \texttt{PTO-test}. As a prototypical ferroelectric material, PbTiO$_3$ is one of the most extensively studied perovskite oxides. Its ground state adopts a tetragonal phase (space group $P4mm$) characterized by spontaneous electrical polarization, which transitions to a nonpolar cubic phase (space group $Pm\bar{3}m$) at temperatures above 760~K, as observed in experiments\cite{Shirane51p265}. The tetragonal phase features a short axis, $a$, and a long axis, $c$, with the tetragonality defined by the ratio $c/a$, which correlates with the magnitude of the polarization.
The energy difference between the ferroelectric and paraelectric phases, determined by DFT calculations at zero Kelvin, is 16 meV/atom.
This moderate energy difference falls within the accuracy range of typical force fields. Furthermore, the sub-800K transition temperature allows for direct MD validation without requiring extrapolation to extreme thermal regimes ($>$1000~K), where anharmonic effects could introduce significant complexities.
For these reasons, we consider the \texttt{PTO-test} an ideal benchmark: it is sufficiently complex to reveal potential limitations of foundational atomistic models in simulating structural dynamics, yet tractable enough to enable systematic error analysis.

The \note{foundational atomistic models} selected for this benchmark include CHGNet, GPTFF, MACE, M3GNet, ORB, and SevenNet. Unfortunately, our request for access to EquiformerV2-OMat\cite{Liao23parXiv,Barroso24parXivv1} was denied. Table~\ref{tab:ml_models_comparison} provides an overview of these MLFFs, detailing the model versions used in our tests, their training datasets, the number of trainable parameters, and the mean absolute error for energy and forces during training. Additionally, we evaluate UniPero, a ``professional model" designed as a force field for perovskite oxides, covering 14 metal elements\cite{Wu23p180104}. It mainly follows the architecture of DPA-1\cite{Zhang24p94}, an earlier version of DPA-2.

\note{\section{Results}}

\begin{figure}
	\begin{center}
		\includegraphics[width=0.6\textwidth]{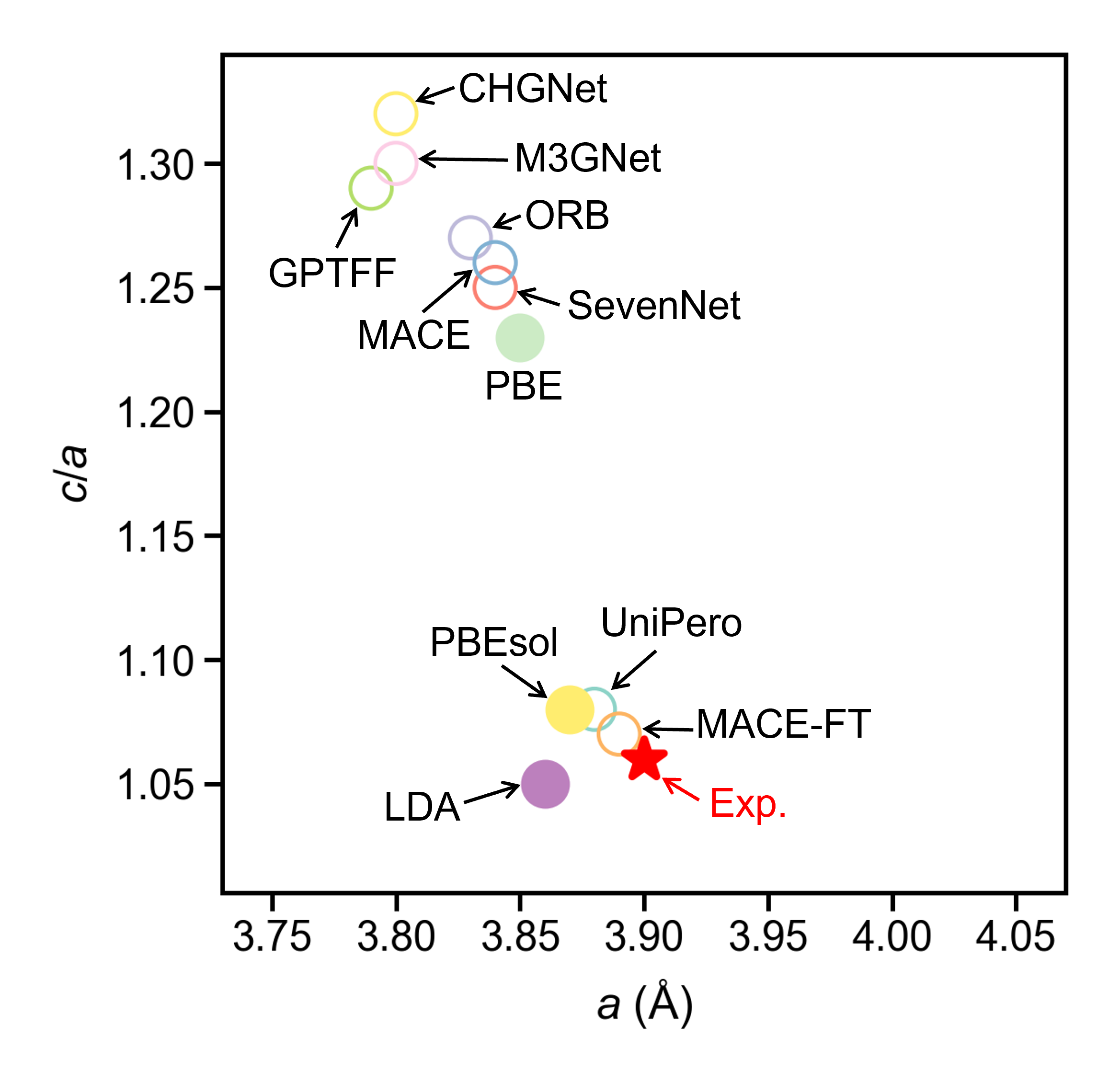}
	\end{center}
	\caption{Lattice parameter $a$ and tetragonality ($c/a$) of ground-state PbTiO$_3$ predicted by various MLFFs and exchange-correlation functionals.  
 } 
 \label{fig:lat}
\end{figure}

\note{\subsection{Static Properties}}
We begin by determining the ground-state structure of the tetragonal phase of PTO through structural optimizations employing various models. Figure~\ref{fig:lat} compares the lattice parameter $a$ and the tetragonality ($c/a$) predicted by these models with results from standard exchange-correlation functionals, including LDA, PBE, and PBEsol. The values are also summarized in Table~\ref{tab:0k}. 
It is well known that the PBE functional significantly overestimates the $c/a$ ratio (experimental value: 1.06), yielding a value of 1.23, whereas PBEsol gives a closer estimate of 1.10. This discrepancy explains why models trained on PBE-based databases, such as CHGNet, M3GNet, and MACE, inherit this bias, predicting $c/a$ ratios even larger than that from PBE itself. The exception is UniPero, which aligns with PBEsol due to its training on PBEsol-derived data. This reveals an expected limitation in foundational atomic models: their accuracy is inherently tied to the exchange-correlation functional used in their training database. For systems like PTO, where even conventional exchange-correlation functionals struggle to reproduce key properties like tetragonality, selecting an appropriate functional \textit{a priori} becomes essential for developing reliable MLFFs. 

To show how such limitations might be addressed, we investigate the effect of fine-tuning. Specifically, we fine-tuned the MACE model on the small PBEsol-based dataset from the UniPero study to create a new model, MACE-FT. This relatively simple procedure appears to be highly effective; as shown in Figure~\ref{fig:lat}, the MACE-FT model yields a ground-state structure in strong agreement with the PBEsol reference, a result the original model did not achieve.

\begin{figure}
	\begin{center}
		\includegraphics[width=1.0\textwidth]{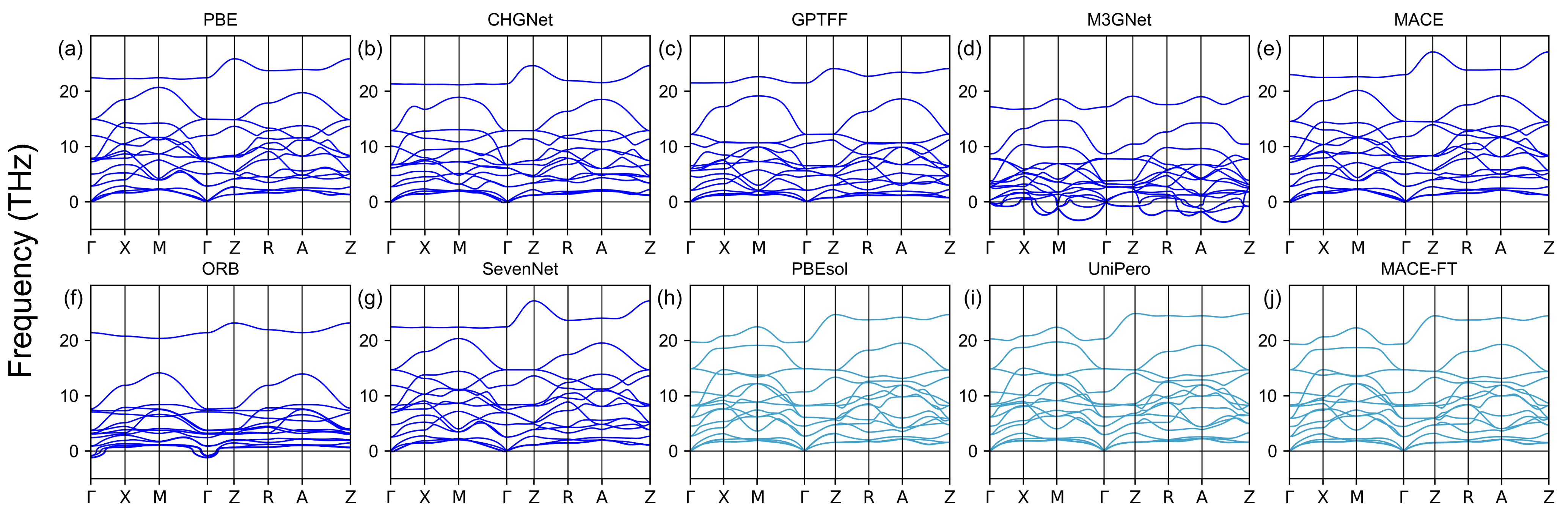}
	\end{center}
	\caption{Phonon spectra of PbTiO$_3$ calculated using various MLFFs, each based on the optimized ground-state tetragonal structure. \notes{The panels show results for: (a) PBE, (b) CHGNet, (c) GPTFF, (d) M3GNet, (e) MACE, (f) ORB, (g) SevenNet, (h) PBEsol, (i) UniPero, and (j) MACE-FT.} The spectra obtained from \notes{(a)} PBE and \notes{(h)} PBEsol are also included for comparison. \notes{(i)} UniPero and \notes{(j)} MACE-FT are trained on a PBEsol-derived database. 
 } 
 \label{fig:phonon}
\end{figure}

We further calculate the phonon spectrum of the optimized tetragonal PTO for each model using the finite-displacement method implemented in Phonopy package\cite{Togo23p353001},  with atomic forces evaluated directly by the respective model. 
Despite their overestimated tetragonality, most models including CHGNet, MACE, and SevenNet generate phonon spectra free of imaginary frequencies (Figure~\ref{fig:phonon}), confirming dynamical stability. As shown in Figure~\ref{fig:phonon}, the phonon spectra of CHGNet, GPTFF, MACE, and SevenNet closely align with the PBE reference. In contrast, the phonon spectrum of M3GNet exhibits instability across the Brillouin zone; ORB displays localized instabilities near the $\Gamma$ point, characterized by weak imaginary frequencies (below 20 cm$^{-1}$), and also predicts notably flat bands for low-frequency phonons. 
Both UniPero and MACE-FT accurately reproduce the PBEsol phonon spectrum. 
Since phonon spectra are highly sensitive to the second derivatives of the potential energy surface near equilibrium, this benchmark highlights that most models effectively capture the local curvature of the energy landscape corresponding to their parent exchange-correlation functional.

\note{\subsection{Finite-Temperature Properties}}

\begin{figure}
	\begin{center}
		\includegraphics[width=1.0\textwidth]{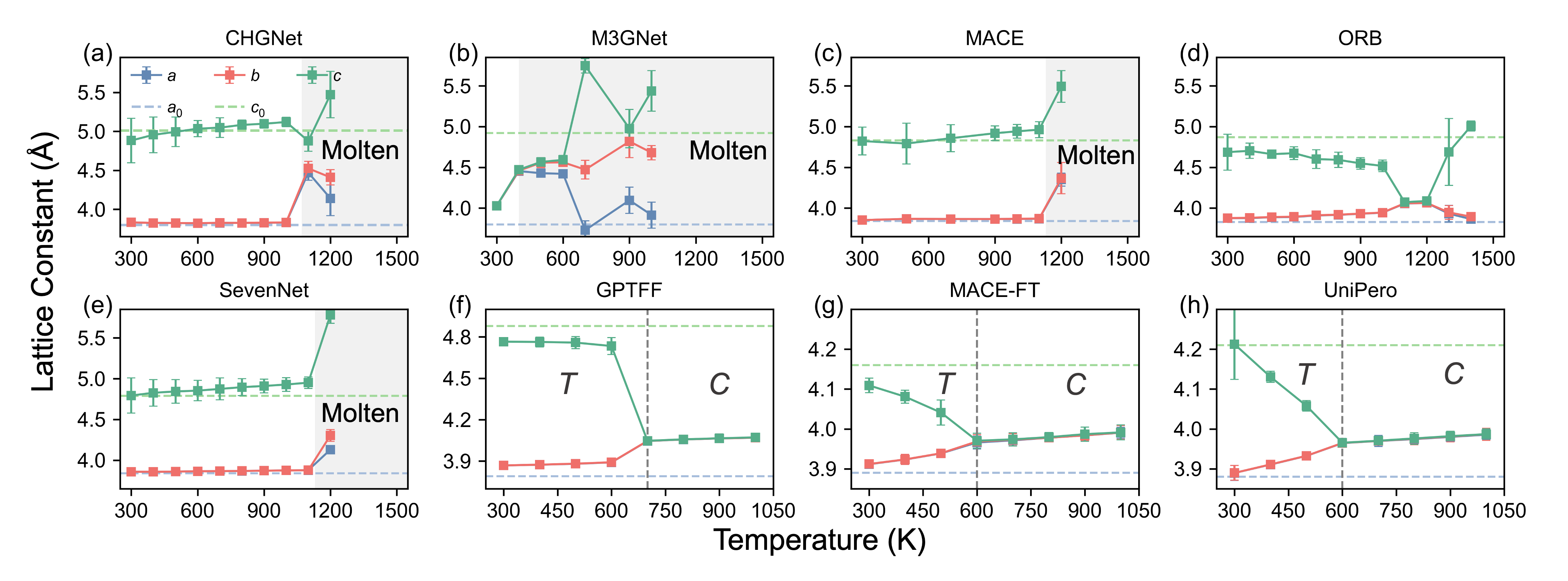}
	\end{center}
	\caption{Temperature-dependent lattice constants ($a$ and $c$) obtained from $NPT$ MD simulations using various machine learning force fields. \notes{The panels show results for: (a) CHGNet, (b) M3GNet, (c) MACE, (d) ORB, (e) SevenNet, (f) GPTFF, (g) MACE-FT, and (h) UniPero.} The dashed lines indicate the ground-state lattice parameters ($a_0$ and $c_0$) of tetragonal PbTiO$_3$ for each force field. \note{The error bars represent the standard deviation over the 50-ps production trajectory, reflecting the extent of thermal fluctuations.}
 } 
 \label{fig:NPT}
\end{figure}

One might expect that an accurate representation of the local potential energy surface near the ground-state structure would ensure at least qualitative reliability for finite-temperature lattice dynamics. However, our findings reveal a significant limitation: most foundational atomistic models struggle to capture dynamic behavior accurately. As shown in Figure~\ref{fig:NPT}, the majority of tested models fail to reproduce the expected temperature-driven tetragonal-to-cubic phase transition during constant-pressure, constant-temperature ($NPT$) MD simulations.  

For example, MD simulations using CHGNet, M3GNet, MACE, and SevenNet show abrupt instabilities above a critical temperature, with the system collapsing into a disordered, molten state. Before melting, CHGNet, MACE, and SevenNet stabilize an unphysical, persistent supertetragonal phase. ORB correctly captures the tetragonal-to-cubic transition near 1100 K but incorrectly predicts a reverse cubic-to-tetragonal transition at higher temperatures. Among the models tested, only GPTFF predicts a temperature-driven (super)tetragonal-to-cubic phase transition.
These results indicate a key limitation: accurate modeling of local curvature near the ground state does not guarantee correct treatment of anharmonic interactions or free-energy landscapes governing temperature-dependent structural transitions. In contrast, both UniPero and MACE-FT successfully reproduce the expected ferroelectric–paraelectric transition, though with an underestimated Curie temperature by approximately 160 K compared to experiment.

\begin{figure}
	\begin{center}
		\includegraphics[width=0.6\textwidth]{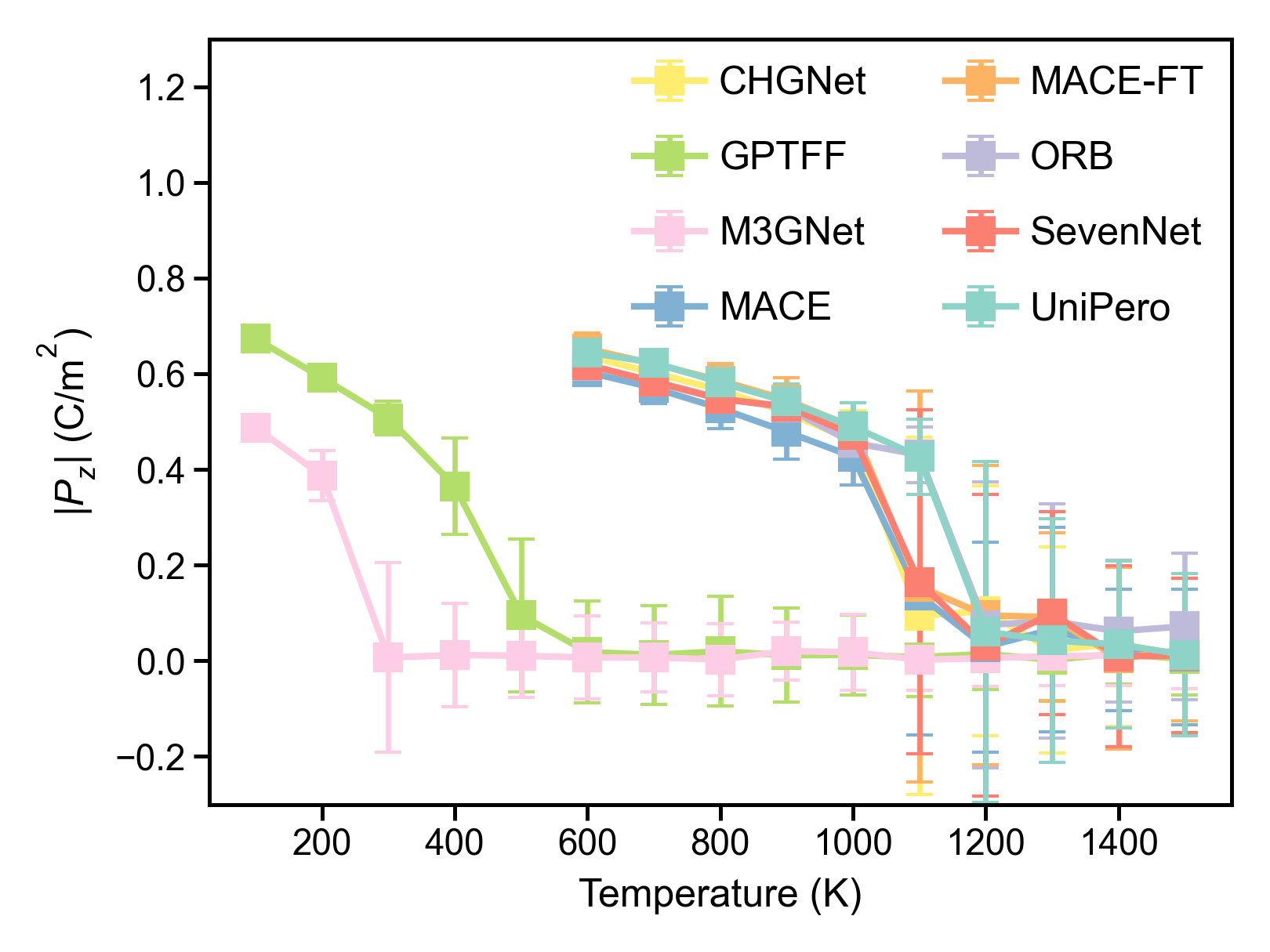}
	\end{center}
	\caption{Temperature-dependent spontaneous polarization along the $c$-axis ($P_z$) obtained from $NVT$ MD simulations, with lattice parameters fixed to the experimental room-temperature values ($a = 3.90$~\AA, $c = 4.15$~\AA). At high temperatures, the polarization does not fully converge to zero due to the imposed tetragonality constraint ($c/a = 1.06$). \note{The error bars represent the extent of polarization fluctuations arising from thermal effects over the 50-ps production trajectory.}
 } 
 \label{fig:NVT}
\end{figure}

Running $NPT$ simulations imposes stringent accuracy requirements on force fields, demanding precise parameterization to capture pressure-density relationships and accurate computation of virial contributions essential for pressure control. However, if a foundational atomistic model is primarily trained on equilibrium configurations (ground-state structures), it may lack the generalizability to handle the dynamic volume fluctuations inherent to $NPT$ ensembles.  To mitigate this challenge, we conduct a controlled validation test using constant-volume, constant-temperature ($NVT$) MD simulations, fixing the lattice constants of PTO to experimental values. This approach eliminates volume relaxation, simplifying the system while still allowing us to probe temperature-driven phase transitions.  
Notably, most MLFFs including CHGNet, MACE, MACE-FT, ORB, SevenNet, and UniPero successfully predict the ferroelectric-to-paraelectric transition, with the spontaneous polarization along the long axis ($P_z$) dropping to near zero at $\approx$1100 K (Fig.~\ref{fig:NVT}). In contrast, M3GNet and GPTFF exhibit significant deviations, predicting Curie temperatures far below expectations.
These results indicate that under constrained $NVT$ conditions, most models capture finite-temperature lattice behavior, as the reduced degrees of freedom simplify the energy landscape. 

\note{\subsection{Computational Efficiency}}

\begin{figure}
	\begin{center}
		\includegraphics[width=0.6\textwidth]{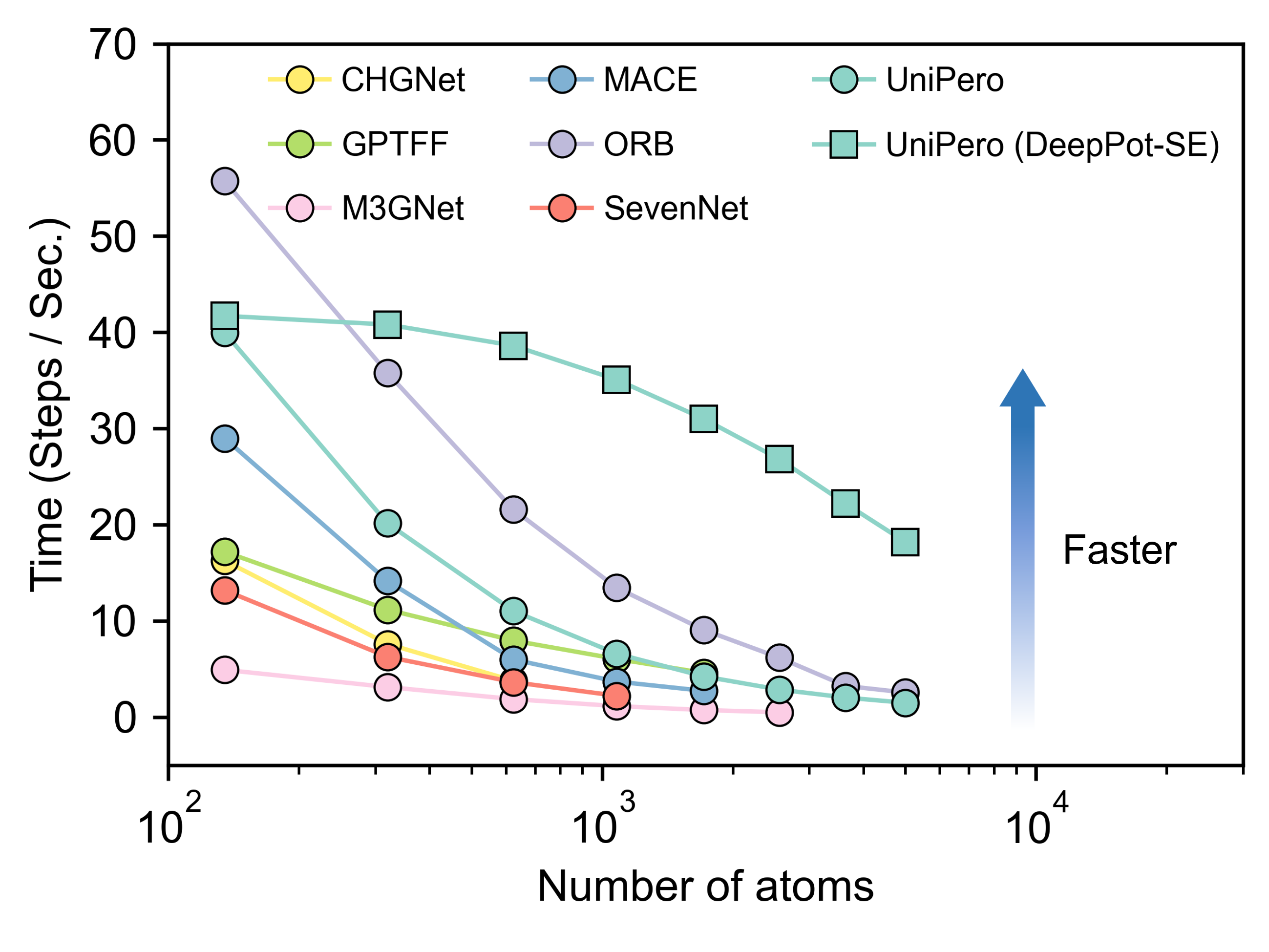}       
	\end{center}
	\caption{Computational efficiency benchmark. The reported speed data are for reference only, as a model's optimal performance can be further improved through careful tuning and the implementation of multi-GPU parallelism.
 } 
 \label{fig:speed}
\end{figure}

\note{While previous reviews have primarily focused on the accuracy of foundational atomistic models, a practitioner must also consider computational efficiency, an aspect that is often equally important, if not more so. In real-world R\&D environments, where computational resources and physical time are limited, the ability to perform large-scale MD simulations efficiently becomes a critical factor in model selection.}
We also briefly discuss the computational efficiency of the tested models, as shown in Figure~\ref{fig:speed}. It is noted that only SevenNet and the DPA-based UniPero are explicitly designed for multi-GPU parallelism, a crucial feature for large-scale MD simulations.
Since some models do not support multi-GPU parallelism or specialized MD packages like LAMMPS~\cite{Plimpton95p1}, the speed test is conducted on a single GPU using the Atomic Simulation Environment (ASE)~\cite{Larsen17p273002}, which integrates each MLFF as a calculator. Therefore, the reported speed data are for reference only, as the optimal performance of a model could be improved with careful tuning.

Our benchmark reveals that most models have yet to fully optimize their performance for GPU acceleration.  For instance, M3GNet’s slower computational performance arises from unresolved GPU compatibility issues in our cluster which defaults to CPU execution rather than leveraging GPU acceleration. While this issue might be resolved with proper settings, it highlights a potential engineering burden for users adopting a foundational atomistic model at this stage.
Notably, ORB outperforms UniPero despite having a larger parameter count, leveraging the TensorFloat32 data format for enhanced efficiency; \note{its non-conservative architecture, which directly predicts forces rather than deriving them via energy gradients, further accelerates computation.} UniPero, originally based on the DPA-1 architecture with the self-attention mechanism, demonstrated significantly improved speed when simplified to a smooth edition of deep potential (DeepPot-SE)~\cite{Zhang18p4441v1} and further compressed using techniques including tabulated inference, operator merging, and precise neighbor indexing, making it the fastest model in our benchmarks. 
Furthermore, by integrating this optimized model into LAMMPS and fully harnessing multi-GPU parallelism, we successfully conducted an MD simulation of 240,000 atoms across 48 GPUs, achieving a computational speed of approximately 42 steps per second.

\note{
\section{Discussion}

While our evaluation centers on a single system, PbTiO$_3$, it was chosen randomly and without tuning to highlight model strengths or weaknesses. The fact that several state-of-the-art foundational atomistic models fail to reproduce well-established finite-temperature behavior in this prototypical material suggests that the issues we reveal are unlikely to be isolated. Instead, they point to broader, systemic challenges that may affect the practical deployment of foundational atomistic models across a wide range of materials and applications. In the following, we highlight three critical areas of concern from a practitioner's perspective: training data quality, computational efficiency and scalability, and the user experience within the current software ecosystem.

\subsection{Training Data Quality}
One of the most critical factors limiting model reliability is the quality of the training data. For MLFFs, this quality hinges on two key aspects: the fidelity of the reference calculations and the diversity of the sampled configurations. The choice of exchange-correlation functional plays a central role in determining reference accuracy. Recent efforts, such as MatPES and others~\cite{kaplan25arxiv,levine25parxiv,Barroso24parXivv1}, have improved upon earlier datasets by adopting more advanced functionals like r$^2$SCAN and sampling configurations from finite-temperature MD simulations. However, our PbTiO$_3$ results underscore that simply switching to a more sophisticated functional does not guarantee better outcomes. For example, a model trained on SCAN still overestimated the $c/a$ ratio and Curie temperature, performing worse than one trained on PBEsol~\cite{xie25p094113}. This highlights that the ``best" exchange-correlation functional can vary by system and often requires prior domain knowledge. 

Fine-tuning offers a practical solution. When we fine-tuned a modern architecture like MACE using a small amount of system-specific data, its performance improved significantly, capturing the correct phase transition. But this introduces a tradeoff: if foundational atomistic models require new DFT data and system-specific tuning to work reliably, their benefits over traditional, from-scratch approaches become less compelling. Therefore, we urge the community to initiate a more rigorous discussion around the ``scaling law" for foundational atomistic models. Specifically, we need quantitative metrics to compare the true cost and speed-up of fine-tuning a large pre-trained model versus training a smaller model from scratch for a given system.

\subsection{Efficiency and Scalability}
Computational efficiency and scalability present another major challenge. Real-world MD simulations often demand long timescales and large system sizes, scenarios where the performance of many foundational atomistic models degrades significantly. This is partially due to their large model sizes, which increase inference cost. Promising strategies are emerging. For example, pretraining-distillation frameworks use a large model to label system-specific data for a smaller, faster model, offering a practical balance between accuracy and speed~\cite{Zhang24p293v1}. 
Interestingly, the NEP89 achieves empirical-potential-like speed while maintaining competitive accuracy across 89 elements by combining an efficient architecture with an innovative data strategy~\cite{liang25parxiv}.

In addition, software engineering plays a critical role. Many graph-neural-network-based MLFFs lack robust multi-GPU parallelism, limiting their scalability on modern hardware. Models like SevenNet, which implement spatial decomposition for multi-GPU execution, represent important progress~\cite{Park24p4857}. We also found that model performance can vary significantly depending on the MD engine used; switching from ASE to LAMMPS yielded substantial speedups in some cases. These observations highlight that realizing the full potential of MLFFs often requires careful optimization across both model architecture and software infrastructure.

\subsection{User Experience}
Finally, the user experience remains a considerable barrier to adoption. Practitioners, especially those outside the ML community, often face steep learning curves when deploying new foundational atomistic models. The current ecosystem is fragmented, with each model packaged in its own framework, requiring users to learn different APIs, data formats, and dependencies. Moreover, integration with professional MD engines like LAMMPS is frequently incomplete or inefficient, making it difficult to run production-level simulations or interface with established analysis tools. Encouragingly, several recent efforts aim to unify and simplify this landscape. The DeePMD-GNN plugin supports external graph neural network models like NequIP and MACE within the DeepMD-kit framework, while DeePMD-kit v3 supports multiple machine learning backends and optimized multi-GPU inference~\cite{zeng25p3154,zeng25p4375}. Similarly, the Materials Graph Library consolidates various GNN-based MLFFs into a single, extensible library~\cite{ko25p253}. These developments represent meaningful progress, but much work remains to create an ecosystem where foundational atomistic models are as easy to use, and as robust, as traditional force fields.
}

\section{Conclusion}
There is little doubt that we are entering an era where AI is transforming scientific inquiry, particularly in computational materials science. \note{While this Perspective does not attempt to chart the entire frontier, it addresses a focused, practical question: Can we trust foundational atomistic models for finite-temperature MD simulations?} Based on our targeted benchmark, the answer is a cautious ``yes", provided their limitations are carefully considered. Many foundational atomistic models demonstrate remarkable accuracy in predicting phonon spectra and equilibrium properties. However, their behavior under realistic MD conditions can be inconsistent, especially in capturing dynamic phase transitions. \note{While our study focuses on a single material, the observed issues likely reflect broader challenges rooted in training data quality, functional choices, and model generalizability.
These problems are not intrinsic flaws of the models themselves, but rather symptoms of a still-maturing ecosystem. Moving forward, we see value in hybrid strategies that combine pretraining with targeted fine-tuning, alongside greater emphasis on benchmarking dynamic performance, improving software integration, and fostering community-driven standards.} Ultimately, the goal is not to replace specialized models, but to develop robust, adaptable tools that extend the reach of atomistic simulations, unlocking AI’s full potential in computational materials discovery.

\note{
\section{Code Availability}
All implementation scripts and some model weights are publicly available at \url{https://github.com/MoseyQAQ/PTO-test}
}

\notes{
\section{Biographies}

\textbf{Denan Li} Denan Li earned his B.S. in Materials Science and Engineering from Ningbo University in 2024. He is currently a Ph.D. candidate in Physics at Westlake University, working in the Multiscale Materials Modeling Laboratory under the supervision of Prof. Shi Liu. His research focuses on the development and application of machine-learning-assisted multiscale methods to understand the dynamic properties of ferroelectrics, with a particular emphasis on organic–inorganic hybrid systems.

\textbf{Jiyuan Yang} Jiyuan Yang earned his B.S. in Physics from Nanjing Normal University in 2017. He received his Ph.D. in Physics from Zhejiang University in 2025 through a joint program with Westlake University. He is currently a postdoctoral fellow at Westlake University, working in Prof. Shi Liu's lab. His research focuses on the application of deep potential molecular dynamics to study ferroelectric domain dynamics. 

\textbf{Xiangkai Chen} Xiangkai Chen received his Ph.D. in Materials Science and Engineering from the University of Science and Technology Beijing in 2023. He is currently a postdoctoral fellow at Westlake University, working in Prof. Shi Liu's lab. His research focuses on the study of oxide plasticity using the deep potential molecular dynamics method.

\textbf{Lintao Yu} Lintao Yu earned his B.S. in Chemistry from Northeast Petroleum University in 2023. He is currently a Ph.D. candidate in Physics at Westlake University in Prof. Shi Liu's lab. His research focuses on the design and construction of high-performance computing infrastructure for large-scale simulations. 

\textbf{Shi Liu} Shi Liu received his B.S. from the University of Science and Technology of China. He obtained his Ph.D. from the University of Pennsylvania in 2015. He continued his postdoctoral research at the Carnegie Institution for Science in Washington, D.C., and later at the Army Research Laboratory. In June 2019, he joined the Department of Physics at Westlake University, where he is now a tenured Associate Professor. His research interests include novel ferroelectrics, emergent topological phases in condensed matter physics, and deep-learning-based large-scale modeling of complex systems. 

}

\section{Acknowledgments}
We acknowledge the supports from National Natural Science Foundation of China (92370104) and Westlake Education Foundation. The computational resource is provided by Westlake HPC Center.

\printbibliography

\clearpage
\newpage

\begin{table}[h]
\caption{Summary of key properties of various MLFFs used for \texttt{PTO-test}.
}
\centering
\footnotesize
\setlength{\tabcolsep}{3pt}
\renewcommand{\arraystretch}{1.1}
\begin{tabular*}{\textwidth}{@{\extracolsep{\fill}}*{7}{c}}
\hline
Model & Version & Training Set (Size)& Parameters& \makecell{Energy MAE\\(meV/atom)} & \makecell{Force MAE\\(meV/\AA)} \\
\hline
CHGNet\cite{Deng23p1031} & 0.3.0 & MPtrj (146K) & 413K & 26 & 60 \\
GPTFF\cite{Xie24p3525} & v2 & Atomly (37.8M) & 502K & 32 & 71 \\
M3GNet\cite{Chen22p718} & MP-2021.2.8-PES & MPF-2021.2.8 (62.8K)& 228K & 18.7 & 63\\
MACE\cite{Batatia23parXivv1} & MP-0b-medium & MPtrj (146K) & 4.69M & 20 & 45 \\
ORB\cite{Neumann24parXivv1} & orb-v2 & MPtrj (146K) + Alexandria (3.1M) & 25.2M & / & /\\
SevenNet\cite{Park24p4857} & 7net-l3i5 & MPtrj (146K) & 1.17M & 8.3 & 29 \\
UniPero\cite{Wu23p180104} & v1 & Customized (19K)& $\approx$500K & 1.75 & 54 \\
\hline
\end{tabular*}

\vspace{2mm}
\parbox{\textwidth}{
\footnotesize 
Notes: 
Some models have been updated since their initial publication. When available, the latest version is used, and the energy and force mean absolute errors (MAEs, if reported) are taken from the latest version if available; otherwise, they are taken from the original references.
Abbreviations: MPtrj = Materials Project trajectories; MPF = Materials Project structure relaxations
}
\label{tab:ml_models_comparison}
\end{table}

\clearpage
\newpage
\begin{table}[h]
\caption{Lattice parameters ($a$ and $c$), tetragonality ($c/a$) of the tetragonal PbTiO$_3$ phase, and the energy difference ($\Delta E$) between the tetragonal and cubic phases predicted by different MLFFs. Experimental and DFT results are also included for comparison.}
\centering
\setlength{\tabcolsep}{4pt} 
\renewcommand{\arraystretch}{1.2} 
\begin{tabular*}{\textwidth}{@{\extracolsep{\fill}}*{5}{c}}
\hline
Model & $a~(\mathrm{\AA})$ & $c~(\mathrm{\AA})$ & $c/a$ & $\Delta E$ (eV/f.u.)\\
\hline
 CHGNet  & 3.80  & 5.01  & 1.32  & 0.24 \\
 GPTFF & 3.79 & 4.88 & 1.29  & 0.25 \\
 M3GNet & 3.80 & 4.92 & 1.30 & 0.16 \\
 MACE & 3.84 & 4.83 & 1.26 & 0.19 \\
 ORB & 3.83 & 4.87 & 1.27 & 0.21 \\
 SevenNet & 3.84 & 4.79 & 1.25  &  0.21 \\
 MACE-FT & 3.89 & 4.16 & 1.07 & 0.08 \\
 UniPero & 3.88 & 4.21 & 1.08 & 0.09 \\
\hline
LDA\cite{Meyer02p104111} & 3.86 & 4.04 & 1.05 	\\
 PBE & 3.85 & 4.73 & 1.23 & 0.20 \\
  PBEsol & 3.87 & 4.20 & 1.08 & 0.08 \\
 \hline
 Exp.\cite{Mabud79p49}  & 3.90 & 4.15  & 1.06\\
\hline
\end{tabular*}
\label{tab:0k}
\end{table}

\clearpage
\newpage

\clearpage
\newpage

\clearpage
\newpage

\clearpage
\newpage

\clearpage
\newpage

\clearpage
\newpage

\textbf{TOC Graphic}
\begin{figure}[htp!]
	\begin{center}
		\includegraphics[width=0.7\textwidth]{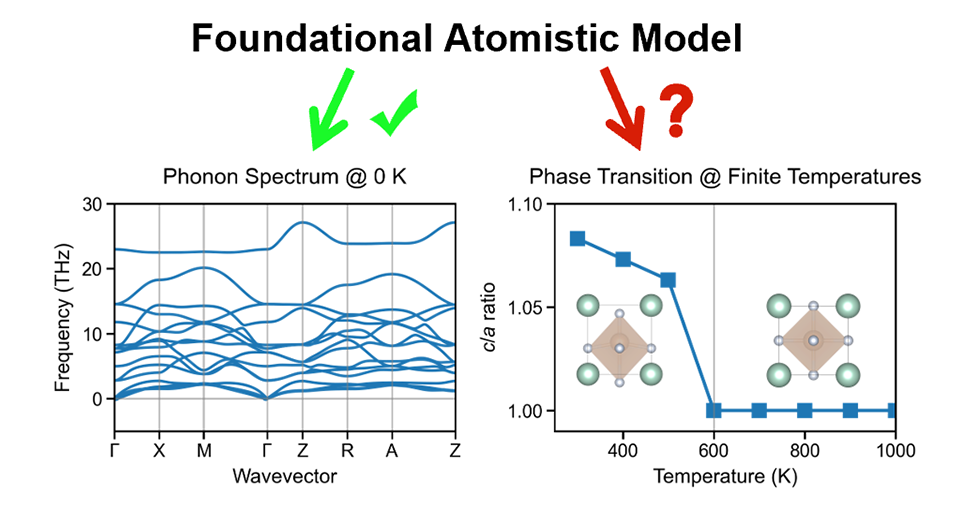}
	\end{center}
    \label{TOC Graphic}
\end{figure}

\end{document}